\documentclass{iaus}
\usepackage{graphicx}

\title[short title of paper] 
{Stellar populations in the center of barred spiral galaxies}

\author[Moll\'{a} et al.]   
{M. Moll\'{a}$^1$, S. Cantin $^2$, C. Robert $^{2}$  \and A. Pellerin $^3$}

\affiliation{$^1$Dpto. de Investigaci\'{o}n B\'{a}sica, CIEMAT, \\ [\affilskip]
$^2$ Universit\'{e} Laval, D\'{e}partement de
physique, de g\'{e}nie physique et d'optique, and Observatoire du mont
M\'{e}gantic, Qu\'{e}bec, G1K 7P4 QC, CANADA  \\ [\affilskip]
$^3$ Space Telescope Science Institute, 3700 San Martin Drive,
MD 21218, Baltimore}

\pubyear{2007}
\volume{245}  
\pagerange{119--126}
\date{?? and in revised form ??}
\jname{Proceedings Title IAU Symposium}
\editors{A.C. Editor, B.D. Editor \& C.E. Editor, eds.}
\begin{document}

\maketitle

\begin{abstract}
We show observations obtained with the integral field spectrometer
OASIS for the central regions of a sample of barred galaxies. The high
spatial resolution of the instrument allows to distinguish various
structures within these regions as defined by stellar populations of
different ages and metallicities.  From these data we obtain important
clues about the star formation history. But we advise that, in order
to obtain adequately the evolutionary sequence, a combination of
chemical and synthesis models may be necessary.

\end{abstract}

\firstsection 

\section{Observing disk galaxies}

When observing a galaxy, we can obtain information about its different
evolutionary phases. On the one hand, we may see emission lines coming
from the gas, which characterize the most recent star forming activity
of the galaxy and provide the nebular chemical abundances. These lines
are usually used as constraints for chemical evolutionary models. On
the other hand, we can measure brightness, colors, and spectral
absorption indices, all of them being stellar population indicators
and giving information about the properties averaged along the whole
time of evolution, and which are interpreted by means of synthesis
models. In this work, we analyze a set of data obtained for the
central regions of a sample of barred galaxies. There are gas and
stars in these regions and, therefore, we will try to constrain the
galaxy history using both types of information.

Our observations were taken with the Canada-France-Hawai Telescope in
2001. We used the instrument OASIS, a microlens matrix imaging
spectrometer in the configuration MR1, to cover the
wavelength range from 4700 to 5500 \AA\, and MR2, from 6200 to
6850 \AA\.  The spatial resolution was 0.42$^{\prime\prime}$ per lenslet 
and the spectral resolution was $\sim$ 2\AA\ per pixel. In this paper,
we show  preliminary results for  4 barred galaxies: NGC 2718
(SABab), NGC 4385 (SB0$+$), NGC 4900 (SBc), and NGC 5430 (SBb).

The data were reduced with the XOASIS package (version 6.0) following
the classical steps: inversion of images, elimination of the
over-scan, subtraction of dark current, subtraction of the bias, mask
extraction, and wavelength calibration. The result was a data-cube
with $\sim$ 800 spectra for each galaxy. Fat field correction,
elimination of cosmic rays, and sky background subtraction were done
on these data-cubes.  We also performed the flux calibration within
IRAF and corrected the spectra for the Galactic extinction using
Seaton's law with the appropriate value of E(B$-$V) for each galaxy
(from NED).  Finally, the spectra were corrected for the galaxy
redshift using NED's velocity and they were binned in order to match
the 0.8$^{\prime\prime}$ spatial resolution limitation of the seeing.

Before the analysis of the data, we measured the intensity of the
emission lines H$\alpha$ and H$\beta$ (taking into account a stellar
absorption component when present) in order to estimate the internal
extinction for each pixel. The lines have been fitted using one or
many gaussians, within IRAF, at the position of the strongest feature
near the theoretical line wavelength values.  The spectra were then
corrected using the reddening law of \cite{car89}.

\section{Results}

The gas oxygen abundances were estimated from the emission line ratios
following the method of \cite{kew02}. The absorption spectral indices
Mg2, Fe5270, Fe5335, and H$\beta$ were measured to obtain the
metallicity and age of the older generation of stars. 

\begin{figure}[!ht]
\includegraphics[width=\textwidth]{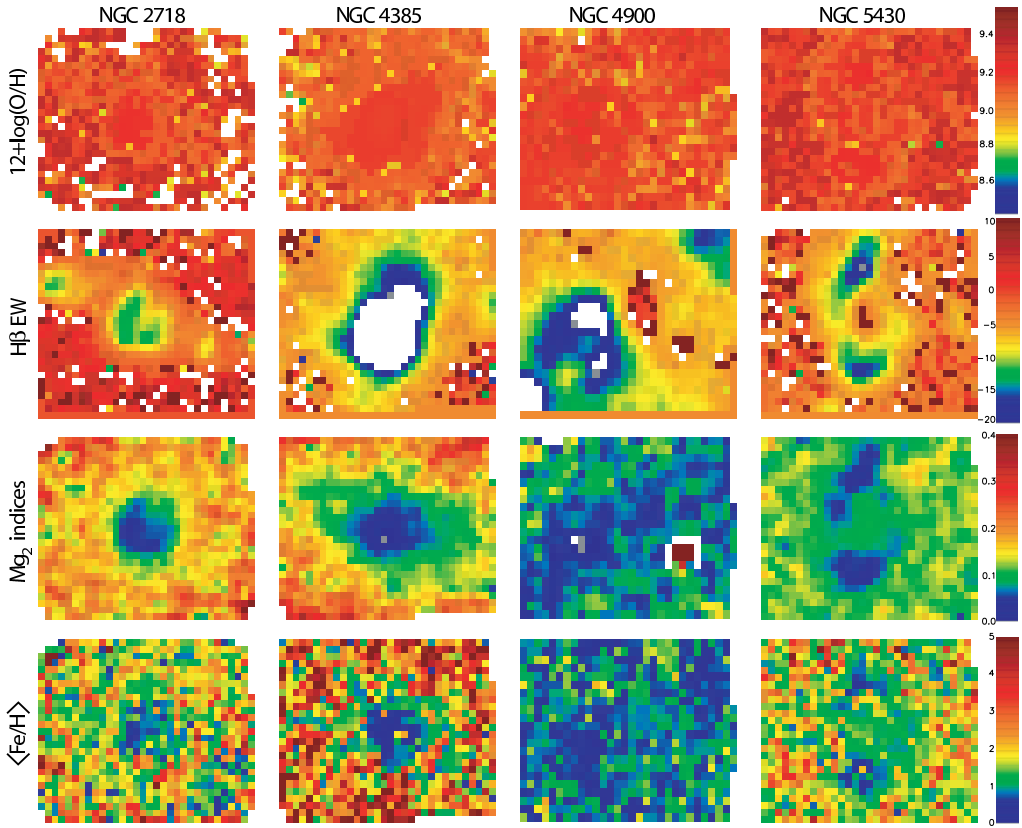}
\caption{Maps shown in rows: 1) oxygen abundance; 2) EW H$\beta$
  (\AA); 3) The index Mg2(mag); 4) The index $\rm <Fe>$ (\AA). Each
  column for a galaxy as labeled. Nort is up and East is left. One
  pixel is 110~pc wide in NGC~2718, 60~pc in NGC~4385, 27~pc in
  NGC~4900, and 82~pc in NGC~5430.}
\label{4gal}
\end{figure}

We show our results in Fig.~\ref{4gal}. In the first row, we see that
oxygen abundances are very uniform within each galaxy, with small
differences between galaxies. In the second row, we show the map of
EW($\rm H\beta$) with positive and negative values, corresponding to
absorption and emission lines.  Each galaxy shows
one (or several) region(s) where $\rm H\beta$ in emission is
intense (dark blue regions), which correspond to the strongest recent
starburst. In NGC~2718, emission is distributed through an elongated  
structure.  In NGC~4385, star formation occurs in the whole central
region while in NGC~4900 the emission is shifted to the South-East.
NGC~5430 shows H{\sc ii} regions distributed around the
center. 

Maps for the spectral indices $\rm Mg2$ and $\rm
<Fe>=(Fe5270+Fe5335)/2$ are in the two bottom rows of Fig.~\ref{4gal}.
Mg2 is more dependent on the age of the stars than $\rm <Fe>$, which
depends strongly on the metallicity Z.  According to these data, Z is
more uniform within each galaxy than the stellar age, which presents
strong differences between the emission regions and the absorption
ones.  In NGC 2718 and NGC 4385, the age maps display a clear gradient
between the inner and the outer regions, the last ones being much
older that the first ones.  In NGC 4900 and NGC~5430, the age seems
more uniform except where strong emission indicates a much younger
population.

\begin{figure}[!ht]
\includegraphics[width=\textwidth]{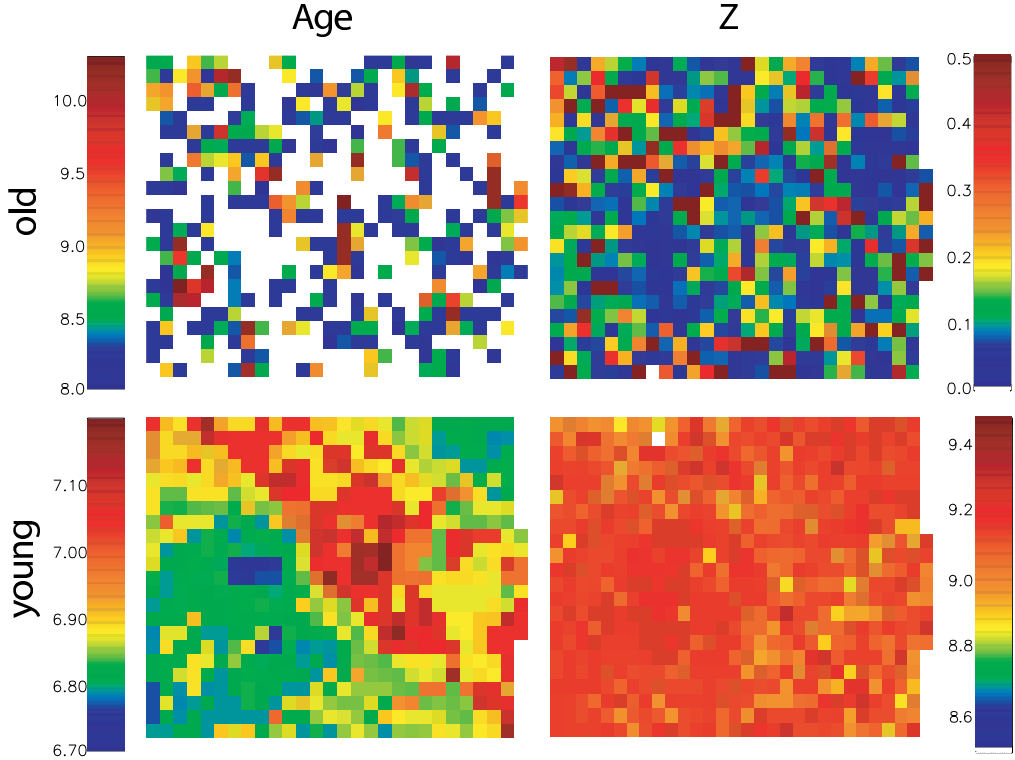}
\caption{The results for NGC~ 4900. Top: age (left) and metallicity Z
(right) for old stellar populations (from  $\rm Mg2$ and $\rm<Fe>$).
Bottom: age and oxygen abundance, $12 + log(O/H)$, for the young
populations (from emission lines).}
\label{ngc4900}
\end{figure}

\section{Interpretation of data: using evolutionary synthesis models}

The emission lines are interpreted using the population synthesis code
LavalSB (\cite{dio06}) to estimate the age of the ionizing young
stellar populations within each region. The absorption spectral
indices are interpreted by means of the evolutionary synthesis models
at high resolution, as it is the case here, from \cite{gon05}.  In
order to estimate the age and Z, we have previously measured the
spectra indices over the model spectra, using the same method as for
the observed spectra. Then, we have performed a Bayesian analysis to
select the best age and metallicity.

The results corresponding to NGC~4900 are in Fig.~\ref{ngc4900} 
The top-right panel shows a map of the stellar metallicity, Z,
which is more or less uniform around $\rm [Fe/H]= Z_{\odot}/3$, with
slight quantitative differences at the North-East where it is higher:
$\rm <[Fe/H]> \simeq 0.6 Z_{\odot}$. The stellar age (top-left panel)
is of few 100 Myr on average, it reaches 1 Gyr as a maximum, with
small and not significant variations among regions.

In the bottom panels of Fig.~\ref{ngc4900}, the age and oxygen
abundance of recent starbursts are shown.  The youngest regions, with
an age around 5-8 Myr, are located at the South-East side and at the
North-West corner. Their average oxygen abundance is high,
$\rm 12+log(O/H) \sim 9.29$, compared to the surrounding.  The
orientation of this structure coincides with the axis of the galaxy
bar. These H{\sc ii} regions may indeed be related to the galaxy bar;
a bar may provoke an inflow of gas from the disk, which results in
starburst regions which, in turn, increases the oxygen abundance.

We have also estimated the stellar masses for the different
components.  On average, we have $\rm < 140x10^{6} M_{\odot}$ for each
30 pc$^{2}$ in the old component. The young burst has a mass of $\rm <
250x10^{3} M_{\odot}$ in the same area, with the highest value along
the galaxy bar. Thus, the mass ratio young/old is near $10^{-3}$.  We
also get some clues about the star formation rates: $\rm SFR_{past}
\simeq 0.014 M_{\odot} yr^{-1} kpc^{-2}$, if we assume that the old
population formation lasted 1~Gyr. If the recent bursts are still
forming stars, then $\rm SFR_{present} \simeq 2.5 \, 10^{-3} - 
1.5 \, 10^{-2} M_{\odot} yr^{-1} kpc^{-2}$.  Thus, the ratio of
past to present star formation rate is around 10 -- 50.

The center of NGC 4900 host, therefore, many star forming
episodes: it suffered a burst in the whole central region 1~Gyr ago 
which may be associated with the bulge population.  Another
burst occurred $\sim$ 100 Myr ago, refilling the very central
region; then 13-14 Myr ago some small knots of formation took place
along the bar structure, with a burst 8 Myr ago at the
North-West and an episode 5-6 Myr ago in the South-East.

Can we interpret these data in terms of an evolutionary sequence? We
might consider two hypotheses: 1) There are only two populations: a
bulge with an old stellar population uniform in age and Z, and then
the bar provoked a abrupt infall of gas which produces the observed
young starbursts.  In this case the classical method of using single
stellar populations (SSPs) to estimate a mean metallicity and age, as
shown before, is valid. 2) If, however, the stars formed in a
posterior phase are the consequence of a slow infall of gas
(Dom\'{\i}nguez-Tenreiro et al., this volume), which is possible given
the differences in ages along the bar, then the situation changes
since a continuous star formation history, $\Psi(t)$, has taken place.
When using SSP's, a given function for $\Psi(t)$ is usually assumed in
those cases (e.g. \cite{moo06,gan07}), and included it in the equation
of deconvolution: $ F_{\lambda}(t)=\int_{0}^{t}
S_{\lambda}(\tau,Z)\Psi(t')dt'$, (where $\tau=t-t'$).  Actually the
SED of each SSP, $S_{\lambda}(\tau,Z)= S_{\lambda}(\tau,Z(t'))$, where
Z(t) varies in time. Therefore, the use of this method may not give
the whole information or the correct conclusions, since the enrichment
history, $\rm Z(t)$, due to the assumed star formation history is not
taken into account.  In order to face this problem, we suggest the use
a set of spectral absorption indices obtained from a chemical
evolutionary model (\cite{mol05}) as templates, to determine
consistently the mean age and metallicity of all the regions.

\end{document}